\documentclass[utf8]{FrontiersinHarvard} 
\setcitestyle{square} 
\usepackage{url,hyperref,lineno,microtype,subcaption}
\usepackage[onehalfspacing]{setspace}
\graphicspath{{figure_eps/}}

\def\keyFont{\fontsize{8}{11}\helveticabold }
\def\firstAuthorLast{Bian {et~al.}} 
	\def\Authors{Xinkai Bian\,$^{1}$, Chaowei Jiang\,$^{1}$ and Xueshang Feng\,$^{1}$}

\def\Address{$^{1}$Institute of Space Science and Applied Technology,
	Harbin Institute of Technology, Shenzhen 518055, China }

\def\corrAuthor{Chaowei Jiang}
\def\corrEmail{chaowei@hit.edu.cn}

\newcommand{\Eq}{{Equation}}

\newcommand{\Fig}{{Figure}}
\newcommand{\Figs}{{Figures}}

\providecommand{\dodoi}[1]{doi:~\href{http://doi.org/#1}{\nolinkurl{#1}}}
\providecommand{\url}[1]{\href{#1}{#1}}
\providecommand{\doeprint}[1]{\href{http://ascl.net/#1}{\nolinkurl{http://ascl.net/#1}}}
\providecommand{\doarXiv}[1]{\href{https://arxiv.org/abs/#1}{\nolinkurl{https://arxiv.org/abs/#1}}}

\begin{document}
\onecolumn
\firstpage{1}

\title[Simulation of solar eruption]{The role of photospheric converging motion in initiation of solar eruptions} 

\author[\firstAuthorLast]{\Authors}
\address{} 
\correspondance{} 

\extraAuth{}

\maketitle

\begin{abstract}
It is well known that major solar eruptions are often produced by active regions with continual photospheric shearing and converging motions. 
Here, through high accuracy magnetohydrodynamics simulation, we show how solar eruption is initiated in a single bipolar configuration as driven by first shearing and then converging motions at the bottom surface. 
Different from many previous simulations, we applied the converging motion without magnetic diffusion, thus it only increases the magnetic gradient across the polarity inversion line but without magnetic flux cancellation. The converging motion at the footpoints of the sheared arcade creates a current sheet in a quasi-static way, and the eruption is triggered by magnetic reconnection of the current sheet, which supports the same scenario as shown in our previous simulation with only shearing motion. 
With the converging motion, the current sheet is formed at a lower height and has a higher current density than with shearing motion alone, which makes reconnection more effective and eruption stronger. Moreover, the converging motion renders a fast decay rate of the overlying field with height and thus favorable for an eruption. 
This demonstrate that the converging flow is more efficient to create the current sheet and more favorable for eruption than by solely the shearing flow.

\tiny
 \keyFont{ \section{Keywords:} Magnetic fields; Magnetohydrodynamics (MHD); Methods: numerical; Sun: corona; Sun: Coronal mass ejections} 
\end{abstract}

\section{Introduction}
Coronal mass ejections (CMEs) are the most spectacular eruptions on the Sun. They represent a explosive release of free magnetic energy stored in the coronal magnetic field. Due to the line-tied effect of the photosphere, the coronal magnetic field is continuously but slowly dragged at their feet by the photospheric surface motions. 
Usually, these motions are organized in large scale, such as shearing along the polarity inversion line (PIL), converging towards the PIL, rotation of sunspot, and dispersion of magnetic flux~\citep{Chintzoglou2019, Brown2003, Min2009, Vemareddy2012, Vemareddy2017, VanDriel-Gesztelyi2003, Lamb2013}.
With these motions (one type of them or their superposition), the coronal magnetic configuration is driven to evolve continually away from the potential field state until a critical point at which no more stable equilibrium can be maintained and an eruption is triggered.
Before the onset of the eruption, the Lorentz force dominates and self-equilibrates in the coronal system, that is, the downward magnetic tension force of the overlying, mostly unsheared flux cancels out the outward magnetic pressure force of the low-lying, strongly stressed flux. 
With initiation of the eruption, the force balance in the coronal system is catastrophically disrupted and the free magnetic energy is rapidly converted into heating (i.e., flare) and acceleration (i.e., CME) of the plasma.

It remains an open question how solar eruptions are initiated and many theories has been proposed~\citep{Forbes2006, Shibata2011, Chen2011a, Schmieder2013, Aulanier2014, Janvier2015}, which generally be classified into two categories, the first based on ideal magnetohydrodynamics (MHD) instabilities and the other on resistive process, i.e., magnetic reconnection. 

But for eruptions initiated from in most common bipolar configuration on the Sun, most models resort to ideal MHD instabilities that require the pre-existence of a magnetic flux rope (MFR) in the corona before eruption. For example, the kink instability and torus instability of the MFR would initiate an eruption~\citep{Kliem2006, Torok2005, Fan2007, Aulanier2010, Amari2018}. 
In the MHD simulations of these models, there needs to be a key phase to transform shear magnetic arcade to MFR before the onset of eruption, such as magnetic cancellation, tether-cutting reconnection, opposite flux emerging~\citep{Amari2000, Amari2003, Amari2003a, Aulanier2010, Zuccarello2015, Kusano2012}.

Recently, through ultra-high resolution MHD simulation, \citet{Jiang2021b} established a simple and effective mechanism of eruption initiation in bipolar field without the need of a pre-existing MFR. Only through a continual surface shearing motion along the PIL, a vertical current sheet (CS) forms quasi-statically, and once the CS is sufficiently thin, fast magnetic reconnection sets in and instantly initiates and subsequently drives the eruption. This scenario is referred to as the BASIC mechanism (here, BASIC is the abbreviation of the key ingredients as involved in the fundamental mechanism: a Bipolar magnetic Arcade as sheared evolves quasi-Statically and forms Internally a Current sheet)~\citep{Bian2022a}. In the simulations of the BASIC mechanism, only a shearing flow is applied to the bipolar while the converging flow has not been considered. Actually the latter has been commonly invoked in flux-cancellation model that builds up unstable MFR~\citep{VanBallegooijen1989, Amari2003a, Zuccarello2015}.

In this paper, we are interested in whether the BASIC mechanism can also be effective if there is converging flow to drive the MHD system other than the shearing flow.
To address this issue, we perform a 3D MHD simulation by first driving the system with shearing motion and then switching to converging motion driving. The shearing driven phase is stopped well before the CS forms.
It is found that the converging motion can still drive the system to form a CS quasi-statically and results in eruption when reconnection sets in, which is essentially the same as when only shearing motion is applied. 
Compared with shearing motion only, the CS is formed at a lower height and has a higher current density, which makes the reconnection more effective and the eruption stronger.
Moreover, the converging motion renders a fast decay rate of the overlying field with height and thus favorable for an eruption. This simulation confirms that the BASIC mechanism also applies with converging flow, and further demonstrates that the converging flow is more efficiently to create the CS and more favorable for eruption than by solely the shearing flow.

This paper is organized as follows. In Sect.~\ref{sec:model}, we show the MHD model along with the initial and boundary conditions, especially the two types of driven motions. Then the simulation results are presented in Sect.~\ref{sec:res}. Finally, our conclusions and discussions are given in Sect.~\ref{sec:con}.

\section{Numerical Model}\label{sec:model}

Our model is a system of full MHD equations in 3D Cartesian coordinates with both plasma pressure and solar gravity, and is solved by the advanced conservation element and solution element method~\citep{Jiang2010, Feng2010, Jiang2016, Jiang2021b}. The basic setup of the model is similar to our previous work \citep{Jiang2021b, Bian2022a}, but here we use two types of driven motions, shear and convergence. 
Our model does not use explicit resistivity in the magnetic induction equation throughout the simulation, but still allows for magnetic reconnection through numerical diffusion when the thickness of the current layer approaches the grid resolution~\citep{Jiang2021b}. By this, we achieved a resistivity as small as we can with a given grid resolution.

The calculation volume is a cube box, which is [$-270, 270$] Mm in the $x$ direction, [$-270, 270$] Mm in the $y$ direction, and [$0, 540$] Mm in the $z$ direction. For convenience, we consider $z=0$ to represent the solar surface (photosphere), since
our model is designed to simulate the coronal evolution as driven by the slow (quasi-statically) surface line-tied motions at the footpoints of the coronal magnetic field lines, and the field line footpoints are anchored at the photosphere.
The volume is large enough such that the simulation runs can be stopped before the disturbance reaches the side and top boundaries.
The full volume is resolved by a block-structured grid with adaptive mesh refinement (AMR) in which the base resolution is $\Delta x=\Delta y=\Delta z=2.88$ Mm, and the highest resolution is $\Delta = 90$ km, which is used to capture the formation process of the CS and the subsequent reconnection. 

\begin{figure*}[htbp!]
\centering
\includegraphics[width=0.8\textwidth]{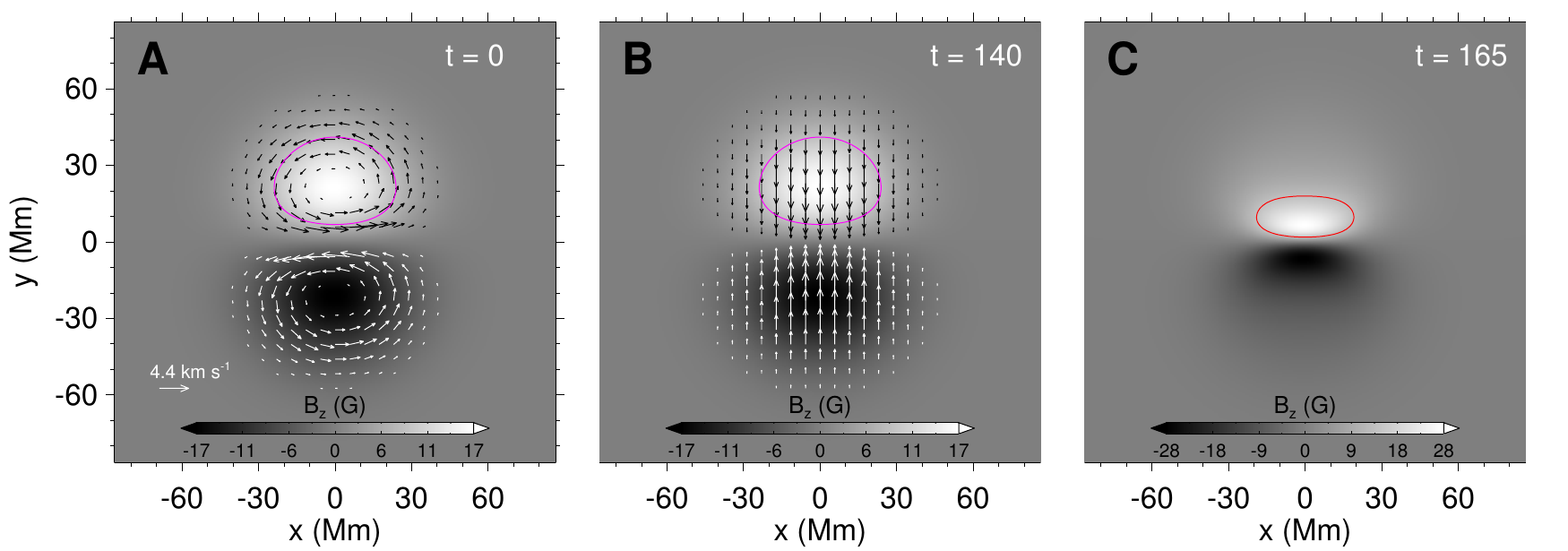}
\caption{The magnetic flux and velocity of the bottom surface during the simulation. \textbf{A}, Magnetic flux distribution and initial surface rotational (shearing) flow (shown by the arrows) at the bottom surface (i.e. $z=0$). \textbf{B}, same as \textbf{A}, but the converging flow starting at $t=140$. \textbf{C}, Magnetic flux distribution at time t=165. The magenta and red lines represent the contours of one half of the maximum of magnetic normal component  (i.e. $\frac{1}{2}B_{z,\rm max}$) in each diagram, respectively.}
\label{fig:initial cond}
\end{figure*}

The initial magnetic field of the MHD simulation is the potential field obtained by the Green's function method from magnetogram. The photospheric magnetogram is a bipolar field modeled by a combination of two Gaussian functions~\citep{Amari2003, Jiang2021b},
\begin{equation}\label{eq:magnetogram}
	\begin{split}
		B_z(x,y,0) = B_0 e^{-x^2/\sigma_x^2}(e^{-(y-y_c)^2/\sigma_y^2}-e^{-(y+y_c)^2/\sigma_y^2}),
	\end{split}
\end{equation} 
where $B_{0}=21.3$ G, $\sigma_{x}=\sigma_{y}=28.8$ Mm, and  $y_c=11.5$ Mm. The parameters $\sigma_{x}$ and $\sigma_{y}$ control the extents of the magnetic flux distribution in the $x$ and $y$ direction, respectively, such a magnetogram is close to two circles (\Fig~\ref{fig:initial cond}A), similar to \citet{Aulanier2010, Zuccarello2015}, but here it is symmetrical.

The initial background atmosphere in the model is in hydrostatic equilibrium, that is, the gravity of the plasma is balanced with the pressure gradient force. Plasma is set to typical coronal values, with a sound speed of $110$ km s$^{-1}$, the maximum Alfv$\acute{\text{e}}$n speed of $1300$ km s$^{-1}$ and the minimum plasma $\beta$ of $1.3\times10^{-2}$. 

Our MHD simulation is first driven by a rotational flows at each magnetic polarity, which is defined as
\begin{equation}\label{eq:rotate}
	\begin{split}
		v_{x}=\dfrac{\partial \psi(B_{z})}{\partial y}; v_{y}=\dfrac{\partial \psi(B_{z})}{\partial x},
	\end{split}
	\end{equation}
with $ \psi $ given by
\begin{equation}\label{eq:rotate_psi}
	\begin{split}
		\psi = v_{0}B_{z}^{2}e^{-(B_{z}^{2}-B_{z, {\rm max}}^{2})/B_{z, {\rm max}}^{2}},
	\end{split}
\end{equation}
where $B_{z,\rm max}$ is the maximum value of $B_{z}$ on the photosphere, and $v_0$ is a constant for scaling such that the maximum of the surface velocity is $4.4$ km s$^{-1}$, which is close to the magnitude of the typical flow speed on the photosphere ($\sim1$ km s$^{-1}$).
The flow speed is smaller than the sound speed by two orders of magnitude and the local Alfv$\acute{\text{e}}$n speed by three orders, respectively, thus representing a quasi-static stress of the coronal magnetic field. 
The rotational flow creates magnetic shear along the PIL and does not change the magnetic flux distribution at the bottom, which is referred to as shearing flow in this paper, and its profile is shown in \Fig~\ref{fig:initial cond}A.
 
The simulation of CASE IV in our previous work~\citep{Bian2022a} was done using the above magnetogram and shearing flow, but the highest gird resolution was $360$ km. Through the continual shearing flow along the PIL, the magnetic energy first increases monotonically over a long period of time, during which the kinetic energy remains at a very low level. Then, at a critical point $t=165$ (the time unit is $\tau=105$ s, all of the times mentioned in this paper are expressed in the same unit), when the thickness of the CS decreases down to the grid resolution, reconnection begins and triggers an eruption, during which the magnetic energy quickly drops and the kinetic energy rises rapidly to nearly $3\%$ of original potential energy. The scenario of this simulation is the BASIC mechanism. 

In order to study whether the BASIC mechanism can also be still effective with converging flow to drive the MHD system other than the shearing flow, we perform this MHD simulation by first driving the system with shearing motion and then switching to converging motion driving, similar to \citet{Amari2003, Zuccarello2015}. 
The shearing driven phase is stopped at $t=140$, a typical moment chosen when the system has accumulated a large amount of free energy but the CS has not yet formed, and then the converging driven phase is started. 
Following \cite{Amari2003}, the converging flow is simply defined as:
\begin{equation}\label{eq:convergenve}
	\begin{split}
		v_x = 0; 
		v_y = -v_{1}B_z(t=0).
	\end{split}
\end{equation}
where $v_1$ is also a constant for scaling such that the largest of the velocity is $4.4$ km s$^{-1}$. 
The converging flow is shown in the \Fig~\ref{fig:initial cond}B. The effect of converging flow on the vertical component of the magnetic field $B_z$ at $t=165$ is shown in the \Fig\ref{fig:initial cond}C. 
The converging flow makes the magnetic flux converge to the PIL, which increases the gradient of the vertical component of the magnetic field $B_z$ along the PIL. 

We do not apply magnetic resistivity in the whole computational volume (including the photosphere surface) during the simulation process, so that magnetic cancellation does not occur. 
The total amount of unsigned flux on the photosphere remains unchanged, while the maximum of the vertical component of the magnetic field $B_z$ becomes larger.

\section{Results}\label{sec:res}
\begin{figure*}[htbp]
	\centering
	\includegraphics[width=0.8\textwidth]{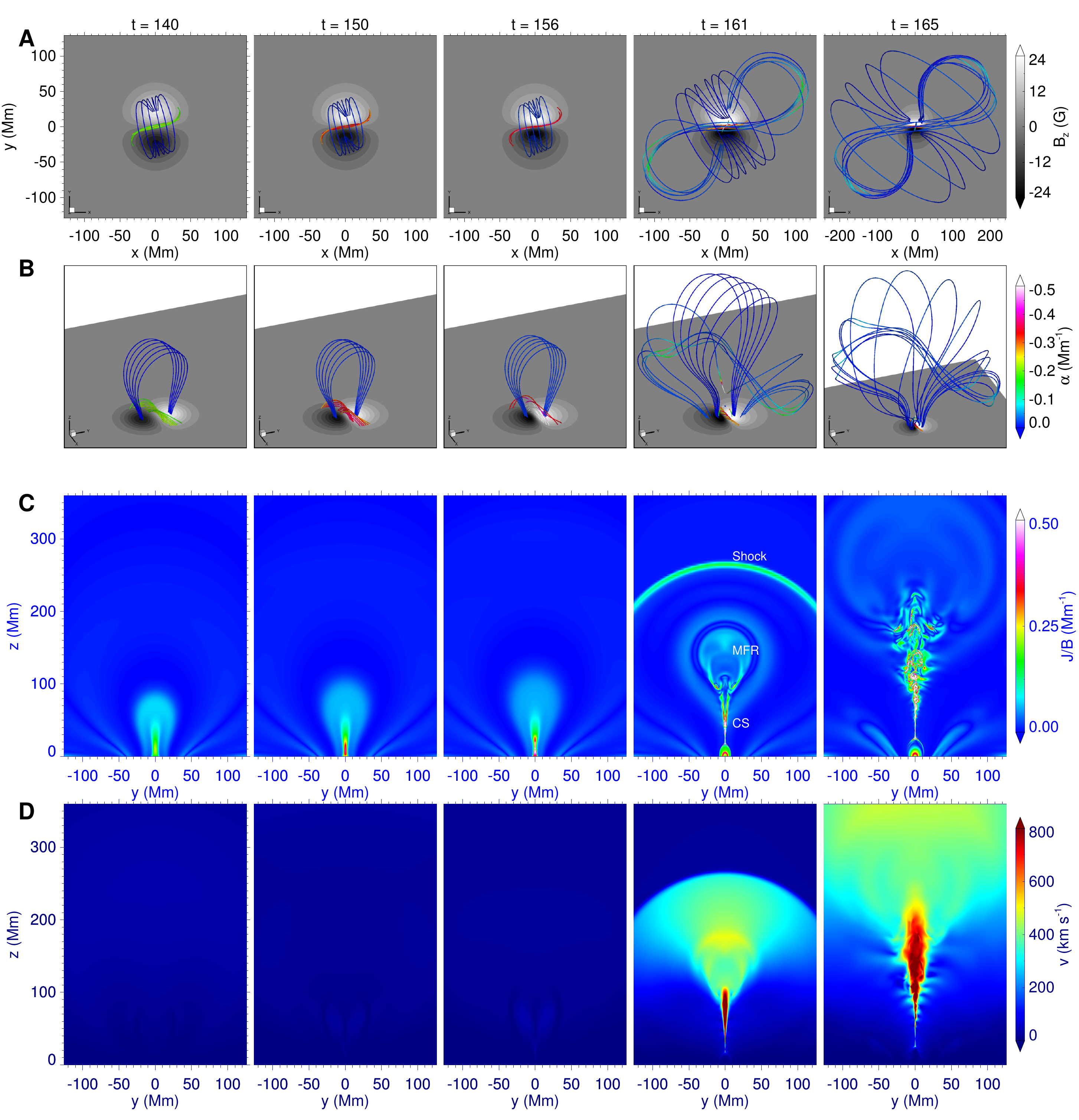}
	\caption{Evolution of magnetic field lines, electric current, and, velocity during simulation process. \textbf{A}, Top view of magnetic field lines. The colored thick lines represent magnetic field line and the colors denote the value of nonlinear force-free factor defined as $\alpha=\textbf{J} \cdot \textbf{B}/B^{2}$, which indicates how much the field lines are non-potential. The background is plotted for the vertical magnetic component $B_z$ on the bottom surface. \textbf{B}, 3D perspective view of the same field lines shown in panel \textbf{A}. The panel with $t=165$ has a larger range. \textbf{C}, Distribution of current density, $J$, normalized by magnetic field strength, $B$, on the vertical cross section, that is, the $x=0$ slice. \textbf{D}, Vertical cross section of the velocity. The maximun velocity and Alfv$\acute{\text{e}}$nic Mach number are also denoted.}
	\label{fig:mag_line}
\end{figure*}

\begin{figure*}[htbp]
	\centering
	\includegraphics[width=0.8\textwidth]{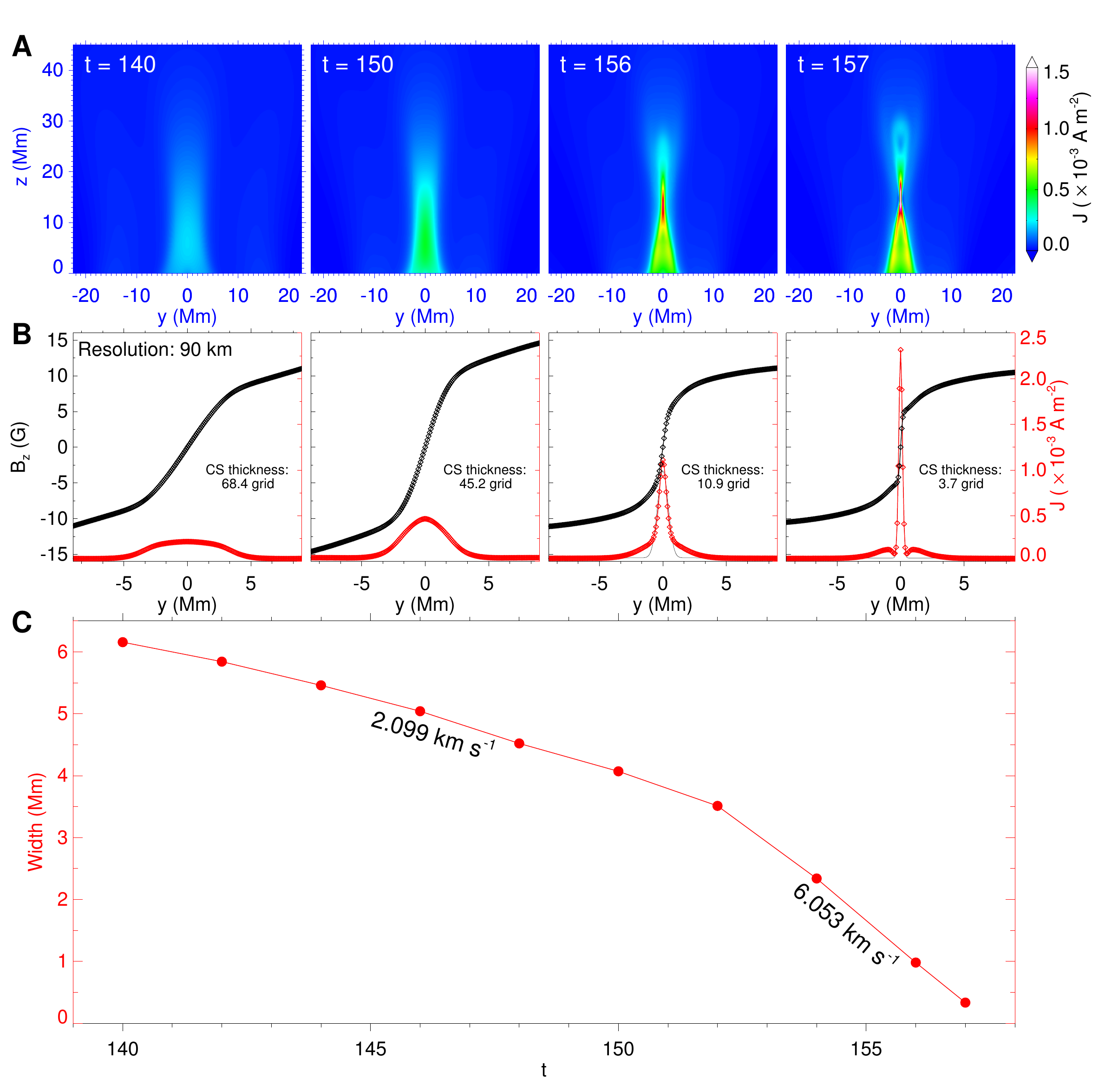}
	\caption{Formation of the CS. \textbf{A}, The distribution of current density on the central vertical slice. \textbf{B}, One-dimensional profile of the vertical component of magnetic field $B_z$ and current density $J$ along a horizontal line crossing perpendicular to the point with the maximum $J$. The diamonds denotes values on the grid nodes. The thickness of CS is denoted, which is defined by the FWHM of a Gauss function fitting (the thin black curve) of the profile of current density. \textbf{C}, Thickness evolution of the CS during converging driven process as shown in \textbf{B}. The velocity at which the thickness of the CS changes is also given. The boundary between these two velocities is $t=152$.}
	\label{fig:Jc_core}
\end{figure*}

Our simulation shows the whole process of the dynamic evolution of the coronal system, including the quasi-static driving process and the subsequent eruption. 
The evolution of magnetic field lines, electric current and velocity during simulation process can be seen from \Fig~\ref{fig:mag_line} and Movie. 
During the shearing driven phase, the evolution of the system is same as our previous work~\citep{Bian2022a}. 
At the end of the shearing driven phase ($t=140$), a magnetic field configuration with strong shear above the PIL is formed, and the current layer formed is still extremely thick (around $70$ grids of highest resolution, see \Fig~\ref{fig:Jc_core}), thus far from a CS.

\begin{figure*}[htbp!]
	\centering
	\includegraphics[width=0.8\textwidth]{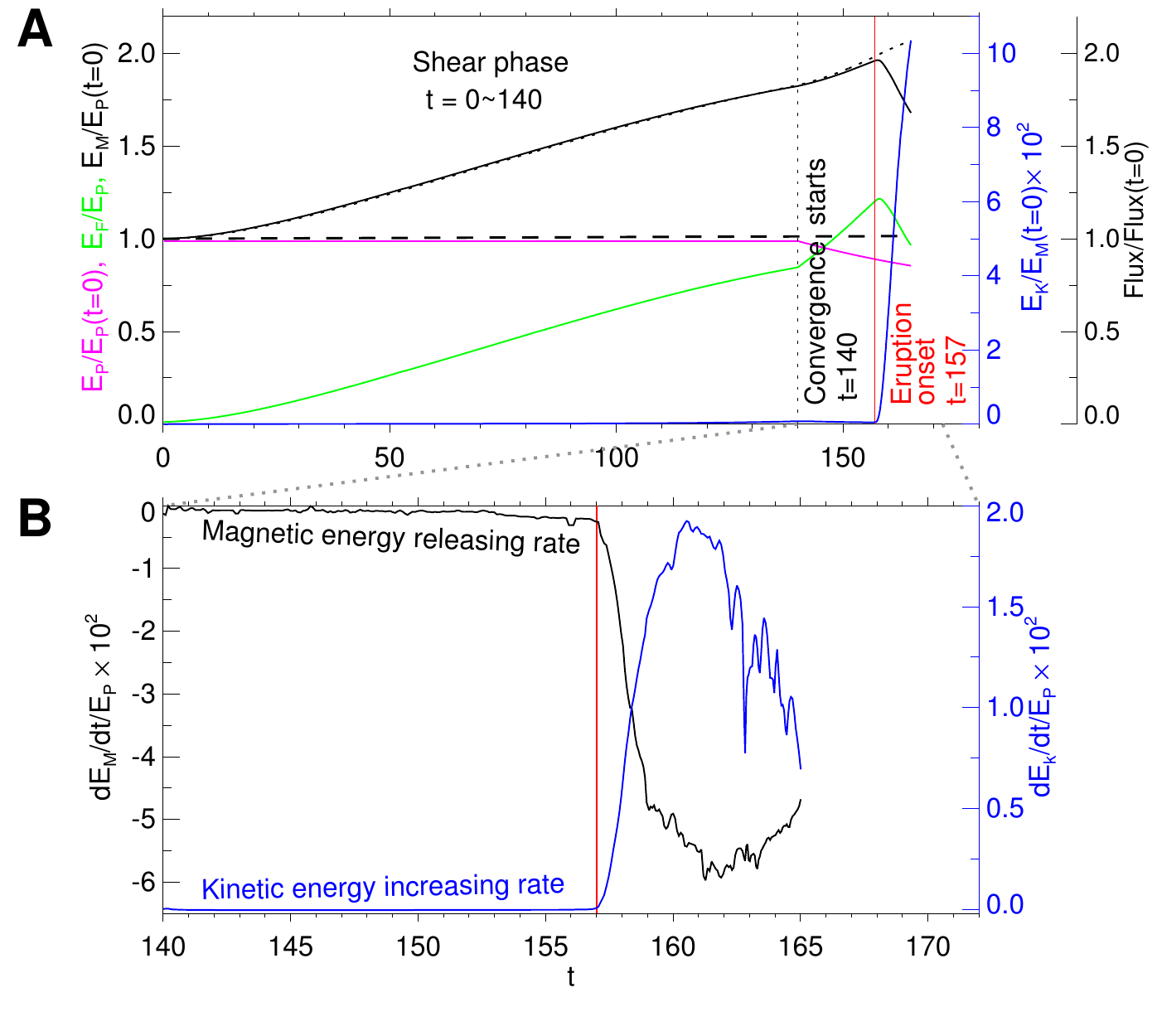}
	\caption{Temporal evolution of magnetic energies, kinetic energy and magnetic flux in the simulation. \textbf{A}, Evolution of total magnetic energy $E_M$ (black), free magnetic energy $E_F$ (green), potential magnetic energy $E_P$ (magenta), kinetic energy $E_K$ (blue line), and magnetic flux(black dashed line). 
	The dotted curve represents the energy injected into the full volume from the bottom boundary through the surface flow.	
	Vertical black dotted line denotes the beginning of the converging flow and vertical red line denotes the transition time from pre-eruption to eruption. 
	\textbf{B}, Magnetic energy releasing rate (black line) and the kinetic energy increasing rate (blue line).}
	\label{fig:energy_evol}
\end{figure*}

At this point, we stop shearing flow and switch on the converging flow, so that the bottom magnetic flux starts to converge to the PIL. Due to the photospheric line-tied effect, the coronal magnetic field is compressed inwards, and the current layer was squeezed into a narrow vertical CS above the PIL (around 3 grids; see \Figs~\ref{fig:mag_line}C and \ref{fig:Jc_core}B). The velocity is negligible, and the total kinetic energy is around $0.03\%$ of the magnetic energy before $t=156$ (\Fig~\ref{fig:energy_evol}A), which indicates that the evolution of the system is quasi-static. 

\Fig~\ref{fig:Jc_core}A shows the structure of the CS in the core region, and the thickness of the CS (at the maximum of current density) is marked in \Fig~\ref{fig:Jc_core}B. It can be seen from \Fig~\ref{fig:Jc_core}C that the thickness of the CS decreases with time at a rate of the order of kilometer per second, which is consistent with converging flow. 
The thickness of the CS decreases more rapidly after $t=152$, about $6.053$ km s$^{-1}$, which is slightly faster than the maximum of converging flow of $4.4$ km s$^{-1}$ applied at the bottom boundary. But the evolution of the system is still quasi-static, compared to the Alfv$\acute{\text{e}}$n speed of thousands of kilometers per second.

\Fig~\ref{fig:energy_evol} presents the evolution of energies during the simulation run from initial time to $t=165$, which includes potential, free and total magnetic energies, as well as kinetic energy.
To compute the potential energy of the system, we use equation $(7)$ in \cite{Amari2003},
\begin{equation}\label{eq:energy_pot}
	\begin{split}
		E_{\rm pot} &= \dfrac{1}{16\pi^{2}} \int_{S\times S'} \dfrac{B_{z}(x,y,0)B_{z}(x',y',0)}{|\textit{\textbf{r}}-\textit{\textbf{r}}'|}{\rm d}s{\rm d}s'.
	\end{split}
\end{equation}
We extract the vertical component of magnetic field $B_z$ on the bottom surface and remap the AMR grid onto a uniform grid of $[400\times400]$ points, and then directly use the \Eq~\ref{eq:energy_pot} to calculate the potential field energy. 
This method is more time efficient than using the potential field extrapolation method~\citep{Bian2022a}. The free magnetic energy is obtained by subtracting the potential energy from the total magnetic energy, which is the volume integral of the magnetic field at a certain moment.
From the beginning of the simulation to $t=140$, during which the system is driven by shearing flow, the potential field energy is constant because the shearing flow does not change the distribution of the bottom surface $B_z$ (\Fig~\ref{fig:energy_evol}). At this stage, the total energy increases monotonically, contributed by the free magnetic energy. 
During the converging phase, i.e., from $t=140$ to the end of simulation, the total energy continues to increase, and the rate of the energy increase is nearly equal to that of the shearing phase~\Fig~\ref{fig:energy_evol}A.
As the converging flows modified the distribution of $B_z$ at the bottom boundary, the potential field energy showing a slightly decrease, so the increase rate of the free magnetic energy is faster.

During the shearing and converging driven phases, the kinetic energy always keep a very low level, confirming that the evolution of the system is quasi-static.
At a critical point ($t=157$), when the thickness of CS reach grid resolution (\Fig~\ref{fig:Jc_core}B), the magnetic reconnection begins and initiates an eruption. The magnetic field energy immediately decreases quickly, meanwhile, the kinetic energy increases rapidly to nearly $10\%$ of original potential energy. 
The beginning of the eruption can also be clearly shown from the time profiles of the magnetic energy release rate and the kinetic energy increase rate, which both have impulsive increases at the onset of the eruption, as shown in \Fig~\ref{fig:energy_evol}B.
This eruption initiated conforms to the BASIC mechanism, that is, the reconnection of the quasi-statically formed CS will immediately trigger the eruption.

As the beginning of magnetic reconnection, a plasmoid (i.e., MFR in $3$D) originates from the tip of the CS and rises rapidly, leaving behind a cusp structure separating the reconnected, post-flare loops from the unreconnected field, as shown in \Fig~\ref{fig:mag_line}C and Movie. The plasmoid expands rapidly, and meanwhile, an arc-shaped fast magnetosonic shock is formed in front of the plasmoid.
Additionally, the turbulence excited by plasmoid-mediated reconnection is also present in our simulations (\Fig~\ref{fig:mag_line}C, at $t\geq161$), which can actually enhance the reconnection rate. 
Therefore, this also shows that the simulation in this paper has very high resolution, as plasmoid-mediated reconnection and the resulted turbulence appear only in simulation with sufficiently high resolutions~\citep{Jiang2021b, Karpen2012} with Lundquist number achieving $\ge 10^5$.

\begin{figure*}[htbp!]
	\centering
	\includegraphics[width=0.6\textwidth]{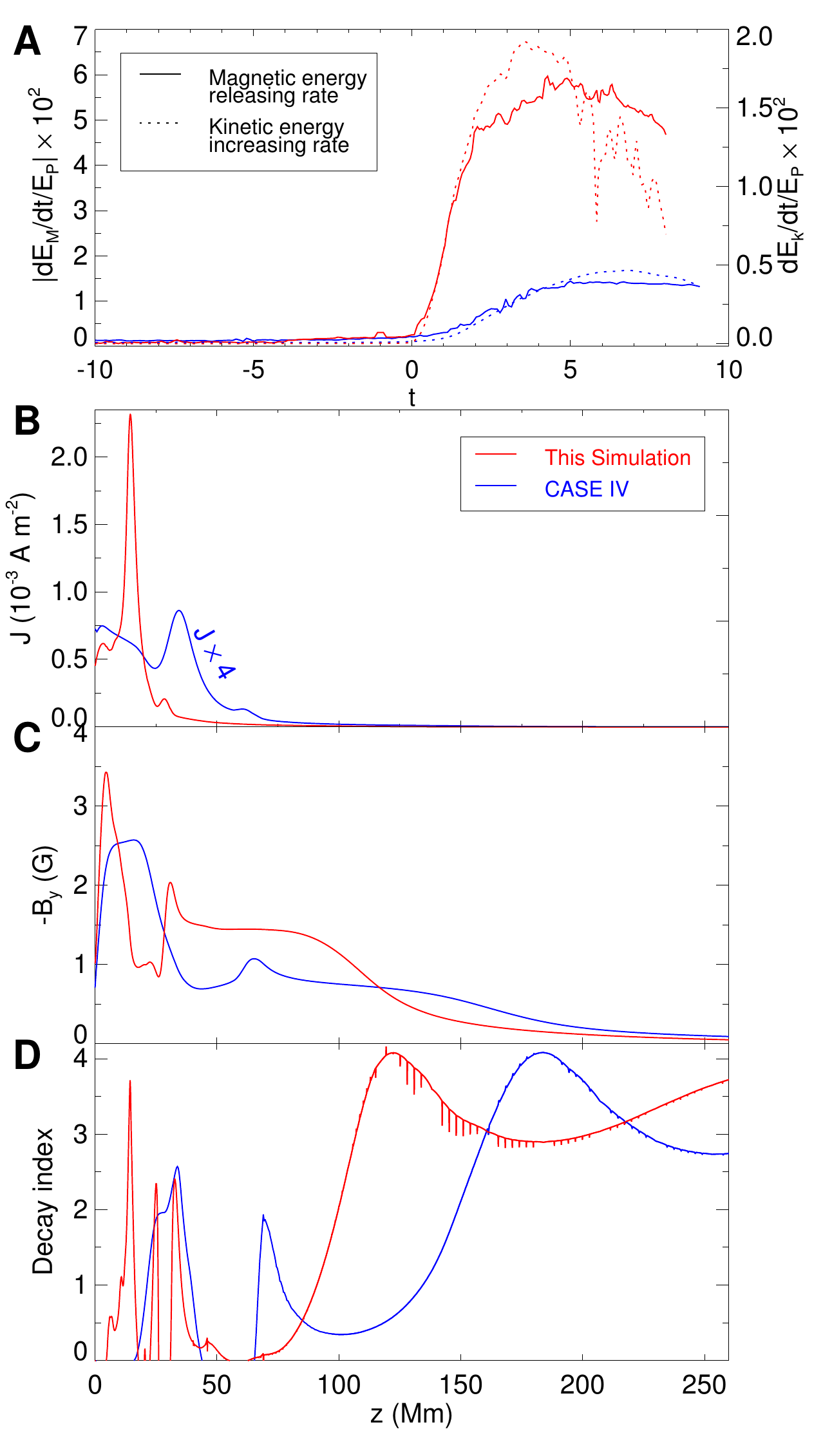}
	\caption{Intensity of eruption, current density, magnetic field and decay index for this eruption and CASE IV.
	\textbf{A}, Magnetic energy releasing rate (solid line) and the kinetic energy increasing rate (dotted line).
	 The timescale is shifted so that the $t=0$ represents the eruption onset. Red and blue curves represent the this simulation and CASE IV, respectively.
	From top to bottom are shown for (\textbf{B}) current density, (\textbf{C}) magnetic field component $B_y$, and (\textbf{D}) decay index of $B_y$, respectively, along $z$ axis, on the eruption onset (this simulation $t=157$ and CASE IV $t=165$).}
	\label{fig:onset_intensity}
\end{figure*}

The intensity of eruption in this simulation was significantly stronger than CASE IV, both in terms of magnetic energy releasing rate and kinetic energy increasing rate, as shown in~\Fig~\ref{fig:onset_intensity}A. Therefore we compared the two simulations, and results are shown in \Fig~\ref{fig:onset_intensity}B-D. \Fig~\ref{fig:onset_intensity}B is the current density distributions of the two simulations along the $z$ axis at the beginning of the eruption. 
The current density of CASE IV is multiplied by $4$, since the highest grid resolution in this simulation is four times that of CASE IV.
The CS in this simulation forms at a lower position and has a larger peak current density, which makes the reconnection more efficient and the eruption stronger.
Moreover, We also check the effect of the background field on the intensity of eruption by calculating the decay index of the background field at the beginning of these two eruptions. 
The overlying field $B_y$ actually plays a key role in constraining the erupting flux rope, which is shown in \Fig~\ref{fig:onset_intensity}C, and the decay index, which is defined as $n=-{\rm d~}{\rm ln}(-B_{y})/{\rm d~}{\rm ln}(z)$, is shown in \Fig~\ref{fig:onset_intensity}D. 
Compared with the CASE IV, the decay index is significantly enhanced in a height of $90$ to $150$ Mm, so that the decay index reaches canonical threshold of 1.5 at lower height, which also favorable for an eruption.

\section{Conclusion}\label{sec:con}
Our simulations demonstrate that the BASIC mechanism is still effective with photospheric driving motion of jointly shearing and converging flows.
At the end of the shearing driven phase, a magnetic field configuration with strong shear above the PIL is formed, and the current layer is still extremely thick. 
With the progress of the converging driven phase, the degree of magnetic shear above the PIL is further strengthened, and the current layer is compressed to be thinner, finally forming a CS and eventually fast magnetic reconnection sets in and instantly initiates the eruption.

During the whole simulation process, the photospheric unsigned flux remains unchanged, that is, magnetic cancellation does not occur. This constraint excludes the possibility of an MFR forms before the onset of eruption, which is the difference between this simulation and \citet{Amari2003a, Zuccarello2015}.

Compared with the CASE IV driven by shearing motion only, the converging motion driven used at $t=140$ advances the formation time of the CS, which is $t=157$ in this simulation compared to $t=165$ in CASE IV, that is, it is more effective for forming CS. In addition,
the CS is formed in this simulation at a lower height and has a higher current density, which makes the magnetic reconnection more effective and the eruption stronger.
Moreover, the converging motion renders a fast decay rate of the overlying field with height and thus favorable for an eruption. 

In summary, this simulation confirms that the BASIC mechanism also applies with converging flow, and further demonstrates that the converging flow is more effective to create the CS and more favorable for eruption than by solely the shearing flow.

%

\section*{Conflict of Interest Statement}

The authors declare that the research was conducted in the absence of any commercial or financial relationships that could be construed as a potential conflict of interest.

\section*{Author Contributions}

XB carried out the simulation and wrote the draft of the manuscript. CJ leads this work and all contribute to the study.

\section*{Acknowledgments}
This work is jointly supported by National Natural Science Foundation of China (NSFC 42174200), the Fundamental Research Funds for the Central Universities (HIT.OCEF.2021033), Shenzhen Science and Technology Program (RCJC20210609104422048), and Shenzhen Technology Project JCYJ20190806142609035. The computational work was carried out on TianHe-1(A), National Supercomputer Center in Tianjin, China.

\bibliographystyle{Frontiers-Harvard} 
\bibliography{mylab}

\begin{thebibliography}{30}
\providecommand{\natexlab}[1]{#1}
\expandafter\ifx\csname urlstyle\endcsname\relax
  \providecommand{\doi}[1]{doi:\discretionary{}{}{}#1}\else
  \providecommand{\doi}{doi:\discretionary{}{}{}\begingroup
  \urlstyle{rm}\Url}\fi
\providecommand{\selectlanguage}[1]{\relax}
\providecommand{\bibAnnoteFile}[1]{%
  \IfFileExists{#1}{\begin{quotation}\noindent\textsc{Key:} #1\\
  \textsc{Annotation:}\ \input{#1}\end{quotation}}{}}
\providecommand{\bibAnnote}[2]{%
  \begin{quotation}\noindent\textsc{Key:} #1\\
  \textsc{Annotation:}\ #2\end{quotation}}

\bibitem[{Amari et~al.(2018)Amari, Canou, Aly, Delyon, and Alauzet}]{Amari2018}
Amari, T., Canou, A., Aly, J.-J., Delyon, F., and Alauzet, F. (2018).
\newblock Magnetic cage and rope as the key for solar eruptions.
\newblock \emph{Nature} 554, 211--215.
\newblock \doi{10.1038/nature24671}
\bibAnnoteFile{Amari2018}

\bibitem[{Amari et~al.(2003{\natexlab{a}})Amari, Luciani, Aly, Mikic, and
  Linker}]{Amari2003}
Amari, T., Luciani, J.~F., Aly, J.~J., Mikic, Z., and Linker, J.
  (2003{\natexlab{a}}).
\newblock Coronal mass ejection: Initiation, magnetic helicity, and flux ropes.
  i. boundary motion\textendash driven evolution.
\newblock \emph{The Astrophysical Journal} 585, 1073--1086.
\newblock \doi{10.1086/345501}
\bibAnnoteFile{Amari2003}

\bibitem[{Amari et~al.(2003{\natexlab{b}})Amari, Luciani, Aly, Mikic, and
  Linker}]{Amari2003a}
Amari, T., Luciani, J.~F., Aly, J.~J., Mikic, Z., and Linker, J.
  (2003{\natexlab{b}}).
\newblock Coronal mass ejection: Initiation, magnetic helicity, and flux ropes.
  ii. turbulent diffusion\textendash driven evolution.
\newblock \emph{The Astrophysical Journal} 595, 1231--1250.
\newblock \doi{10.1086/377444}
\bibAnnoteFile{Amari2003a}

\bibitem[{Amari et~al.(2000)Amari, Luciani, Mikic, and Linker}]{Amari2000}
Amari, T., Luciani, J.~F., Mikic, Z., and Linker, J. (2000).
\newblock A twisted flux rope model for coronal mass ejections and two-ribbon
  flares.
\newblock \emph{The Astrophysical Journal} 529, L49--L52.
\newblock \doi{10.1086/312444}
\bibAnnoteFile{Amari2000}

\bibitem[{Aulanier(2014)}]{Aulanier2014}
Aulanier, G. (2014).
\newblock The physical mechanisms that initiate and drive solar eruptions.
\newblock In \emph{Nature of Prominences and Their Role in Space Weather}. vol.
  300, 184--196.
\newblock \doi{10.1017/S1743921313010958}
\bibAnnoteFile{Aulanier2014}

\bibitem[{Aulanier et~al.(2010)Aulanier, T{\"o}r{\"o}k, D{\'e}moulin, and
  DeLuca}]{Aulanier2010}
Aulanier, G., T{\"o}r{\"o}k, T., D{\'e}moulin, P., and DeLuca, E.~E. (2010).
\newblock Formation of torus-unstable flux ropes and electric currents in
  erupting sigmoids.
\newblock \emph{The Astrophysical Journal} 708, 314--333.
\newblock \doi{10.1088/0004-637X/708/1/314}
\bibAnnoteFile{Aulanier2010}

\bibitem[{Bian et~al.(2022)Bian, Jiang, Feng, Zuo, Wang, and Wang}]{Bian2022a}
Bian, X., Jiang, C., Feng, X., Zuo, P., Wang, Y., and Wang, X. (2022).
\newblock Numerical simulation of a fundamental mechanism of solar eruption
  with a range of magnetic flux distributions.
\newblock \emph{Astronomy \& Astrophysics} 658, A174.
\newblock \doi{10.1051/0004-6361/202141996}
\bibAnnoteFile{Bian2022a}

\bibitem[{Brown et~al.(2003)Brown, Nightingale, Alexander, Schrijver, Metcalf,
  Shine et~al.}]{Brown2003}
Brown, D., Nightingale, R., Alexander, D., Schrijver, C., Metcalf, T., Shine,
  R., et~al. (2003).
\newblock Observations of rotating sunspots from trace.
\newblock \emph{Solar Physics} 216, 79--108.
\newblock \doi{10.1023/A:1026138413791}
\bibAnnoteFile{Brown2003}

\bibitem[{Chen(2011)}]{Chen2011a}
Chen, P.~F. (2011).
\newblock Coronal mass ejections: Models and their observational basis.
\newblock \emph{Living Reviews in Solar Physics} 8.
\newblock \doi{10.12942/lrsp-2011-1}
\bibAnnoteFile{Chen2011a}

\bibitem[{Chintzoglou et~al.(2019)Chintzoglou, Zhang, Cheung, and
  Kazachenko}]{Chintzoglou2019}
Chintzoglou, G., Zhang, J., Cheung, M. C.~M., and Kazachenko, M. (2019).
\newblock The origin of major solar activity: Collisional shearing between
  nonconjugated polarities of multiple bipoles emerging within active regions.
\newblock \emph{The Astrophysical Journal} 871, 67.
\newblock \doi{10.3847/1538-4357/aaef30}
\bibAnnoteFile{Chintzoglou2019}

\bibitem[{Fan and Gibson(2007)}]{Fan2007}
Fan, Y. and Gibson, S.~E. (2007).
\newblock Onset of coronal mass ejections due to loss of confinement of coronal
  flux ropes.
\newblock \emph{The Astrophysical Journal} 668, 1232--1245.
\newblock \doi{10.1086/521335}
\bibAnnoteFile{Fan2007}

\bibitem[{Feng et~al.(2010)Feng, Yang, Xiang, Wu, Zhou, and Zhong}]{Feng2010}
Feng, X., Yang, L., Xiang, C., Wu, S.~T., Zhou, Y., and Zhong, D. (2010).
\newblock Three-dimensional solar wind modeling from the sun to earth by a
  sip-cese mhd model with a six-component grid.
\newblock \emph{The Astrophysical Journal} 723, 300--319.
\newblock \doi{10.1088/0004-637X/723/1/300}
\bibAnnoteFile{Feng2010}

\bibitem[{Forbes et~al.(2006)Forbes, Linker, Chen, Cid, K{\'o}ta, Lee
  et~al.}]{Forbes2006}
Forbes, T.~G., Linker, J.~A., Chen, J., Cid, C., K{\'o}ta, J., Lee, M.~A.,
  et~al. (2006).
\newblock Cme theory and models: Report of working group d.
\newblock \emph{Space Science Reviews} 123, 251--302.
\newblock \doi{10.1007/s11214-006-9019-8}
\bibAnnoteFile{Forbes2006}

\bibitem[{Janvier et~al.(2015)Janvier, Aulanier, and
  D{\'e}moulin}]{Janvier2015}
Janvier, M., Aulanier, G., and D{\'e}moulin, P. (2015).
\newblock From coronal observations to mhd simulations, the building blocks for
  3d models of solar flares (invited review).
\newblock \emph{Solar Physics} 290, 3425--3456.
\newblock \doi{10.1007/s11207-015-0710-3}
\bibAnnoteFile{Janvier2015}

\bibitem[{Jiang et~al.(2021)Jiang, Feng, Liu, Yan, Hu, Moore
  et~al.}]{Jiang2021b}
Jiang, C., Feng, X., Liu, R., Yan, X., Hu, Q., Moore, R.~L., et~al. (2021).
\newblock A fundamental mechanism of solar eruption initiation.
\newblock \emph{Nature Astronomy} 5, 1126--1138.
\newblock \doi{10.1038/s41550-021-01414-z}
\bibAnnoteFile{Jiang2021b}

\bibitem[{Jiang et~al.(2010)Jiang, Feng, Zhang, and Zhong}]{Jiang2010}
Jiang, C., Feng, X., Zhang, J., and Zhong, D. (2010).
\newblock Amr simulations of magnetohydrodynamic problems by the cese method in
  curvilinear coordinates.
\newblock \emph{Solar Physics} 267, 463--491.
\newblock \doi{10.1007/s11207-010-9649-6}
\bibAnnoteFile{Jiang2010}

\bibitem[{Jiang et~al.(2016)Jiang, Wu, Feng, and Hu}]{Jiang2016}
Jiang, C., Wu, S.~T., Feng, X., and Hu, Q. (2016).
\newblock Data-driven magnetohydrodynamic modelling of a flux-emerging active
  region leading to solar eruption.
\newblock \emph{Nature Communications} 7, 11522.
\newblock \doi{10.1038/ncomms11522}
\bibAnnoteFile{Jiang2016}

\bibitem[{Karpen et~al.(2012)Karpen, Antiochos, and DeVore}]{Karpen2012}
Karpen, J.~T., Antiochos, S.~K., and DeVore, C.~R. (2012).
\newblock The mechanisms for the onset and explosive eruption of coronal mass
  ejections and eruptive flares.
\newblock \emph{The Astrophysical Journal} 760, 81.
\newblock \doi{10.1088/0004-637X/760/1/81}
\bibAnnoteFile{Karpen2012}

\bibitem[{Kliem and T{\"o}r{\"o}k(2006)}]{Kliem2006}
Kliem, B. and T{\"o}r{\"o}k, T. (2006).
\newblock Torus instability.
\newblock \emph{Physical Review Letters} 96, 255002.
\newblock \doi{10.1103/PhysRevLett.96.255002}
\bibAnnoteFile{Kliem2006}

\bibitem[{Kusano et~al.(2012)Kusano, Bamba, Yamamoto, Iida, Toriumi, and
  Asai}]{Kusano2012}
Kusano, K., Bamba, Y., Yamamoto, T.~T., Iida, Y., Toriumi, S., and Asai, A.
  (2012).
\newblock Magnetic field structures triggering solar flares and coronal mass
  ejections.
\newblock \emph{The Astrophysical Journal} 760, 31.
\newblock \doi{10.1088/0004-637X/760/1/31}
\bibAnnoteFile{Kusano2012}

\bibitem[{Lamb et~al.(2013)Lamb, Howard, DeForest, Parnell, and
  Welsch}]{Lamb2013}
Lamb, D.~A., Howard, T.~A., DeForest, C.~E., Parnell, C.~E., and Welsch, B.~T.
  (2013).
\newblock Solar magnetic tracking. iv. the death of magnetic features.
\newblock \emph{The Astrophysical Journal} 774, 127.
\newblock \doi{10.1088/0004-637X/774/2/127}
\bibAnnoteFile{Lamb2013}

\bibitem[{Min and Chae(2009)}]{Min2009}
Min, S. and Chae, J. (2009).
\newblock The rotating sunspot in ar 10930.
\newblock \emph{Solar Physics} 258, 203--217.
\newblock \doi{10.1007/s11207-009-9425-7}
\bibAnnoteFile{Min2009}

\bibitem[{Schmieder et~al.(2013)Schmieder, D{\'e}moulin, and
  Aulanier}]{Schmieder2013}
Schmieder, B., D{\'e}moulin, P., and Aulanier, G. (2013).
\newblock Solar filament eruptions and their physical role in triggering
  coronal mass ejections.
\newblock \emph{Advances in Space Research} 51, 1967--1980.
\newblock \doi{10.1016/j.asr.2012.12.026}
\bibAnnoteFile{Schmieder2013}

\bibitem[{Shibata and Magara(2011)}]{Shibata2011}
Shibata, K. and Magara, T. (2011).
\newblock Solar flares: Magnetohydrodynamic processes.
\newblock \emph{Living Reviews in Solar Physics} 8.
\newblock \doi{10.12942/lrsp-2011-6}
\bibAnnoteFile{Shibata2011}

\bibitem[{T{\"o}r{\"o}k and Kliem(2005)}]{Torok2005}
T{\"o}r{\"o}k, T. and Kliem, B. (2005).
\newblock Confined and ejective eruptions of kink-unstable flux ropes.
\newblock \emph{The Astrophysical Journal} 630, L97--L100.
\newblock \doi{10.1086/462412}
\bibAnnoteFile{Torok2005}

\bibitem[{{van Ballegooijen} and Martens(1989)}]{VanBallegooijen1989}
{van Ballegooijen}, A.~A. and Martens, P. C.~H. (1989).
\newblock Formation and eruption of solar prominences.
\newblock \emph{The Astrophysical Journal} 343, 971.
\newblock \doi{10.1086/167766}
\bibAnnoteFile{VanBallegooijen1989}

\bibitem[{{van Driel-Gesztelyi} et~al.(2003){van Driel-Gesztelyi}, Demoulin,
  Mandrini, Harra, and Klimchuk}]{VanDriel-Gesztelyi2003}
{van Driel-Gesztelyi}, L., Demoulin, P., Mandrini, C.~H., Harra, L., and
  Klimchuk, J.~A. (2003).
\newblock The long-term evolution of ar 7978: The scalings of the coronal
  plasma parameters with the mean photospheric magnetic field.
\newblock \emph{The Astrophysical Journal} 586, 579--591.
\newblock \doi{10.1086/367633}
\bibAnnoteFile{VanDriel-Gesztelyi2003}

\bibitem[{Vemareddy(2017)}]{Vemareddy2017}
Vemareddy, P. (2017).
\newblock Successive homologous coronal mass ejections driven by shearing and
  converging motions in solar active region noaa 12371.
\newblock \emph{The Astrophysical Journal} 845, 59.
\newblock \doi{10.3847/1538-4357/aa7ff4}
\bibAnnoteFile{Vemareddy2017}

\bibitem[{Vemareddy et~al.(2012)Vemareddy, Ambastha, and
  Maurya}]{Vemareddy2012}
Vemareddy, P., Ambastha, A., and Maurya, R.~A. (2012).
\newblock On the role of rotating sunspots in the activity of solar active
  region noaa 11158.
\newblock \emph{The Astrophysical Journal} 761, 60.
\newblock \doi{10.1088/0004-637X/761/1/60}
\bibAnnoteFile{Vemareddy2012}

\bibitem[{Zuccarello et~al.(2015)Zuccarello, Aulanier, and
  Gilchrist}]{Zuccarello2015}
Zuccarello, F.~P., Aulanier, G., and Gilchrist, S.~A. (2015).
\newblock Critical decay index at the onset of solar eruptions.
\newblock \emph{The Astrophysical Journal} 814, 126.
\newblock \doi{10.1088/0004-637X/814/2/126}
\bibAnnoteFile{Zuccarello2015}

\end{thebibliography}

\section*{Figure captions}

\begin{figure*}[htbp]
	\centering
	\includegraphics[width=0.8\textwidth]{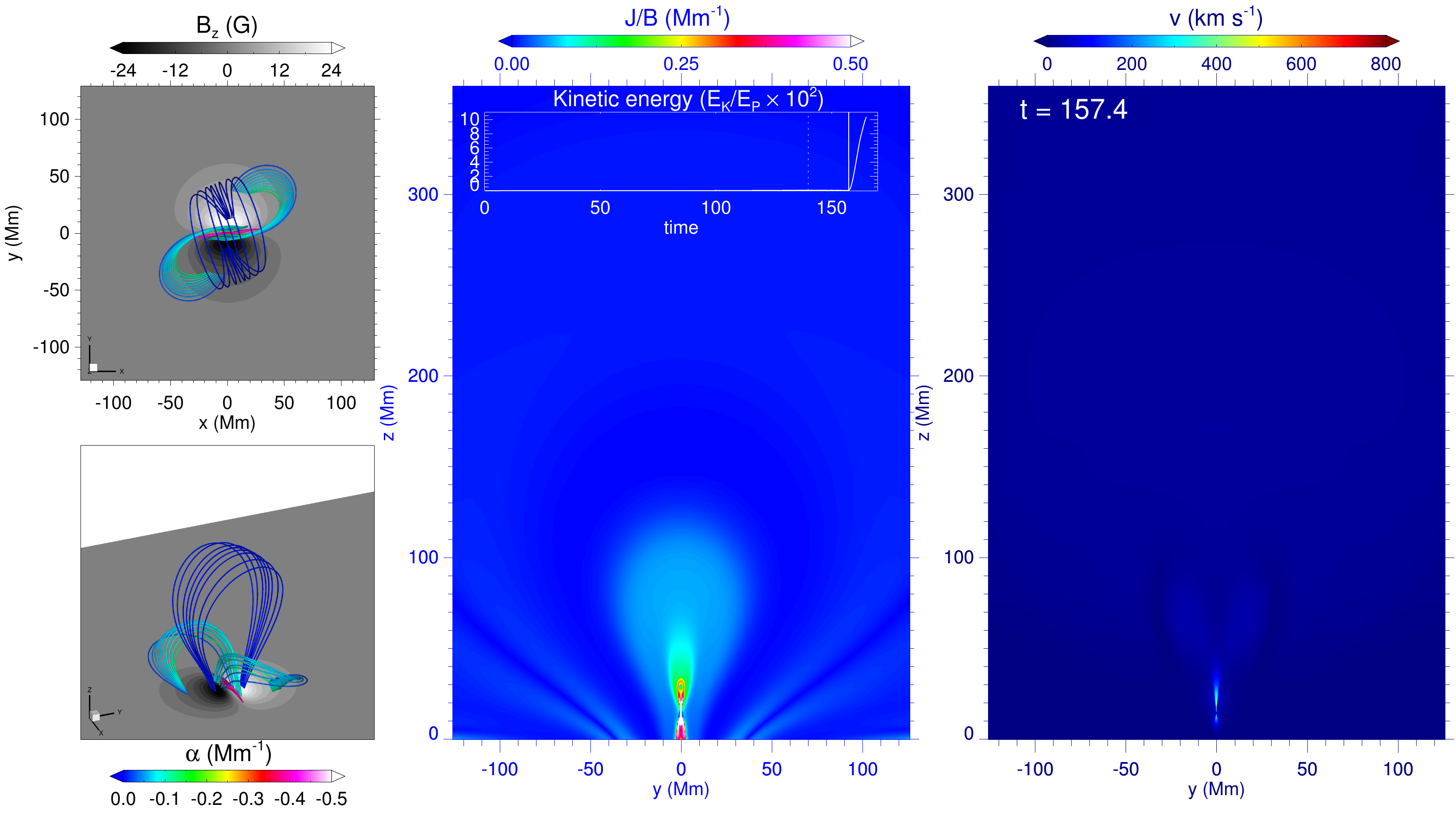}
	\caption{Screenshot of an animation of the \Fig~\ref{fig:mag_line} in the whole simulation process from the initial time to $t=165$. This figure is available as an animation.}
	\label{fig:movie}
\end{figure*}

\end{document}